\documentstyle[buckow,graphicx]{article}

\newcommand{\AmS}{{\protect\the\textfont2
  A\kern-.1667em\lower.5ex\hbox{M}\kern-.125emS}}

\begin {document}

\def\titleline{ 
{\vskip-2.0cm\hfill\small hep-ph/0112269, TIFR/TH/01-50\vskip1.5cm }
Quark number susceptibilities from lattice QCD\footnote{Based on work done 
\cite{qsus,fsus} with Sourendu Gupta and Pushan Majumdar.} 
}

\def\authors{Rajiv V. Gavai}

\def\addresses{
Department of Theoretical Physics \\
Tata Institute of Fundamental Research \\
Homi Bhabha Road, Mumbai 400 005, India \\
{\tt E-mail: gavai@tifr.res.in}
}

\def\abstracttext{ 
Results from our recent investigations of quark number susceptibilities
in both quenched and 2-flavour QCD are presented as a function of valence quark
mass and temperature.  A strong reduction ($\sim$40\%) is seen in the strange
quark susceptibility above $T_c$ in both the cases.  A comparison of our isospin
susceptibility results with the corresponding weak coupling expansion reveals
once again the non-perturbative nature of the plasma up to $3T_c$.  Evidence
relating the susceptibility to another non-perturbative phenomena, 
pionic screening lengths, is presented.}

\large
\makefront

\section{Introduction}

Quantum chromodynamics (QCD) is the theory of interactions of quarks and gluons
which are the basic constituents of strongly interacting particles such as
protons, neutrons and pions.  A complete lack of any experimental observation
of a free quark or gluon led to the hypothesis of their permanent confinement
in the observable particles.  As pointed out by Satz \cite{satz} in his talk,
relativistic heavy ion collision experiments at BNL, New York, and CERN, Geneva
offer the possibility of counting the number of degrees of freedom of strongly
interacting matter at high temperatures, thereby providing a strong argument
for Heisenberg to accept the physical reality of quarks and gluons in spite of
their confinement at lower temperatures.  The basic idea here is to look for
the production of quark-gluon plasma (QGP), predicted by QCD to exist beyond a
transition temperature $T_c$, in the heavy ion collisions and when found,
determine its thermodynamical properties, such as, {\it e.g.}, its energy
density. At very high temperatures, these properties can be computed
theoretically and are seen to be directly proportional to the number of quarks
and gluons.  In order to test this idea, however, one has to face the fact that
the temperatures reached in the current, and near future, experiments are
likely to lie below about 5-10 $T_c$.   The current best theoretical estimates
for thermodynamic observables in this temperature range are provided by
numerical simulations of QCD defined on a discrete space-time lattice
\cite{creutz}.   It seems therefore prudent to evaluate as many independent
observables as possible and compare them with both experiments and approximate
analytical methods.  Quark number susceptibilities constitute a useful
independent set of observables for testing this basic idea of counting the
degrees of freedom of strongly interacting matter.  

Investigations of quark number susceptibilities from first principles can have
direct experimental consequences as well since quark flavours such as electric
charge, strangeness  or baryon number can provide diagnostic tools for the
production of flavourless quark-gluon plasma in the central region of heavy
ion collisions.  It has been pointed out recently \cite{bm,vk} on the basis of
simple models for the hadronic and QGP phases that the fluctuations of such
conserved charges can be very different in these two phases and thus can act as
probes of quark deconfinement.  Indeed, excess strangeness production has been
suggested as a signal of quark-gluon plasma almost two decades ago \cite{jrbm}.
Lattice QCD can provide a very reliable and robust estimate for these
quantities in {\em both} the phases since in thermal equilibrium they are
related to corresponding susceptibilities by the fluctuation-dissipation
theorem:
\begin{equation}
\langle \delta Q^2 \rangle \propto {T \over V} {{\partial^2 \log Z} \over 
\partial \mu_Q^2 } = \chi_Q(T, \mu_Q=0) ~.
\label{flds}
\end{equation}
Here $\mu_Q$ is the chemical potential for a conserved charge $Q$, and $Z$ is
the partition function of strongly interacting matter in volume $V$ at
temperature $T$.  Unfortunately, the fermion determinant in QCD becomes complex
for any nonzero chemical potential for most quantum numbers including those
mentioned above.  Consequently, Lattice QCD is unable to handle finite chemical
potential satisfactorily at present, and cannot thus yield any reliable
estimates of any number density. However, the susceptibility above, i.e., the
first derivative of the number density at zero chemical potential, can be 
obtained reasonably well using conventional simulation techniques, 
facilitating thereby a nontrivial extension of our theoretical knowledge
in the nonzero chemical potential direction.

Quark number susceptibilities also constitute an independent set of observables
to probe whether quark-gluon plasma is weakly coupled in the temperature regime
accessible to the current and future planned heavy ion experiments (say, 1 $
\le  T/T_c \le $ 10).  A lot of the phenomenological analysis of the heavy ion
collisions data is usually carried out assuming a weakly interacting plasma
although many lattice QCD results suggest otherwise.  It has been suggested
\cite{resum} that resummations of the finite temperature perturbation theory
may provide a bridge between phenomenology and the lattice QCD by explaining
the lattice results starting from a few $T_c$.  As we will see below, quark
number susceptibilities can act as a cross-check of the various resummation
schemes.  Earlier work on susceptibilities \cite{rly} did not attempt to
address this issue and were mostly restricted to temperatures very close to
$T_c$.  Furthermore, the quark mass was chosen there to vary with temperature
linearly.  We improve upon them by holding quark mass fixed in physical units
($m/T_c$ = constant). We also cover a larger range of temperature up to 3
$T_c$ and the accepted range of strange quark mass in our simulations.

\section{Formalism}

After integrating the quarks out, the partition function $Z$ for QCD at finite
temperature and density is given by 
\begin{equation}
  Z = \int{\cal D}U {\rm e}^{-S_g}
            \det M(m_u,\mu_u)\det M(m_d,\mu_d)\det M(m_s,\mu_s).
\label{zqcd}
\end{equation}
Here $\{U_\nu(x)\}$, $\nu$ = 0--3, denote the gauge variables and $S_g$ is the
gluon action, taken to be the standard \cite{creutz} Wilson action in our
simulations.  Due to the well-known ``fermion doubling problem'' \cite{creutz},
one has to face the choice of fermion action with either exact chiral symmetry
or violations of flavour symmetry. The Dirac matrices $M$ depend on this
choice.  Since we employ staggered fermions, the matrices, $M$, are of
dimensions 3$N_s^3 N_t$, with $N_s(N_t)$ denoting the number of lattice sites
in spatial(temporal) direction. These fermions preserve some chiral symmetry at
the expense of flavour violation.  Although, they are strictly defined for four
flavours, a prescription exists to employ them for arbitrary number of flavours
which we shall use.  $m_f$ and $\mu_f$ are quark mass and chemical potential
(both in lattice units) for flavour $f$, denoting up(u), down(d), and
strange(s) in eq.(\ref{zqcd}).  The chemical potential needs to be introduced
on lattice as a function $g(\mu)$ and $g(-\mu)$ multiplying the gauge variables
in the positive and negative time directions respectively, such that
\cite{chem} i) $g(\mu) \cdot g(-\mu)$ = 1 and ii) the correct continuum limit
is ensured.  While many such functions $g$ can be constructed, $\exp(\mu)$
being a popular choice, the results for susceptibilities at $\mu=0$ can easily
be shown to be independent of the choice of $g$ even for finite lattice spacing
$a$.  From the $Z$ in eq. (\ref{zqcd}), the quark number densities and the
corresponding susceptibilities are defined as
\begin{equation}
   n_f \equiv \frac{T}{V}\frac{\partial\ln Z}{\partial\mu_f} \qquad
   \chi_{ff'} \equiv \frac{\partial n_f}{\partial \mu_{f'}}
             = \frac{T}{V}
      \left[\frac1Z\frac{\partial^2Z}{\partial\mu_f\partial\mu_{f'}}
          -\frac1Z\frac{\partial Z}{\partial\mu_f}\,
           \frac1Z\frac{\partial Z}{\partial\mu_{f'}}\right]~,
\label{incomplete}
\end{equation}
where the volume $V = N_s^3 a^3$ and the temperature $T = (N_t a)^{-1}$.
To lighten the notation, we shall put
only one subscript on the diagonal parts of $\chi$.
 
In order to obtain information for quark-gluon plasma in the central region,
we evaluate the susceptibilities at the point $\mu_f=0$ for all $f$.  In this
case, each $n_f$ vanishes, a fact that we utilize as a check on our numerical
evaluation.  Moreover, the product of the single derivative terms in eq.\
(\ref{incomplete}) vanishes, since each is proportional to a number density.
We set $m_u=m_d<m_s$.  Noting that staggered quarks have four flavours by
default, $N_f=4$, and defining $\mu_3 = \mu_u - \mu_d$, one finds from eq.
(\ref{incomplete}) that the isotriplet and strangeness susceptibilities are
given by 
\begin{equation}
\chi_3 = {T \over 2V} {\cal O}_1(m_u), \qquad
\chi_s = {T \over 4V} [{\cal O}_1(m_s) + {1 \over 4} {\cal O}_2(m_s)]~,~
\label{susc}
\end{equation}
where ${\cal O}_1 = \langle {\rm Tr} (M''M^{-1} - M'M^{-1}M'M^{-1}) \rangle$,
${\cal O}_2 = \langle ({\rm Tr} M'M^{-1})^2 \rangle$,  
$M'=\partial M/\partial\mu$ and $M''=\partial^2 M/\partial\mu^2$.  
The angular brackets denote averaging with respect to the $Z$ in 
eq. ({\ref{zqcd}).  One can similarly define 
baryon number and charge susceptibilities. We refer the reader for more details
on them to Ref. \cite{fsus}. 

In the discussion above, quark mass appears as an argument of ${\cal O}_i$ and
implicitly in the Boltzmann factor of $Z$.  Let us denote it by $m_{val}$ and
$m_{sea}$ respectively.  While the two should ideally be equal, we evaluated
the expressions above in steps of improving approximations (and increasing
computer costs) by first setting $m_{sea} = \infty $ for all flavours (quenched
approximation \cite {qsus}) and then simulating two light dynamical flavours,
by setting $m_{sea}/T_c = 0.1$ (2-flavour QCD \cite{fsus}).   In each case we
varied $m_{val}$ over a wide range to cover both light u,d quarks as well as
the heavier strange quark.  Details of our simulations as well as the technical
information on how the thermal expectation values of ${\cal O}_i$ were 
evaluated are in Refs. \cite{qsus,fsus}.

\section{Results}

\begin{figure}[htb]
\vspace{-0.5cm}
\begin{center}
\includegraphics*[width=28pc]{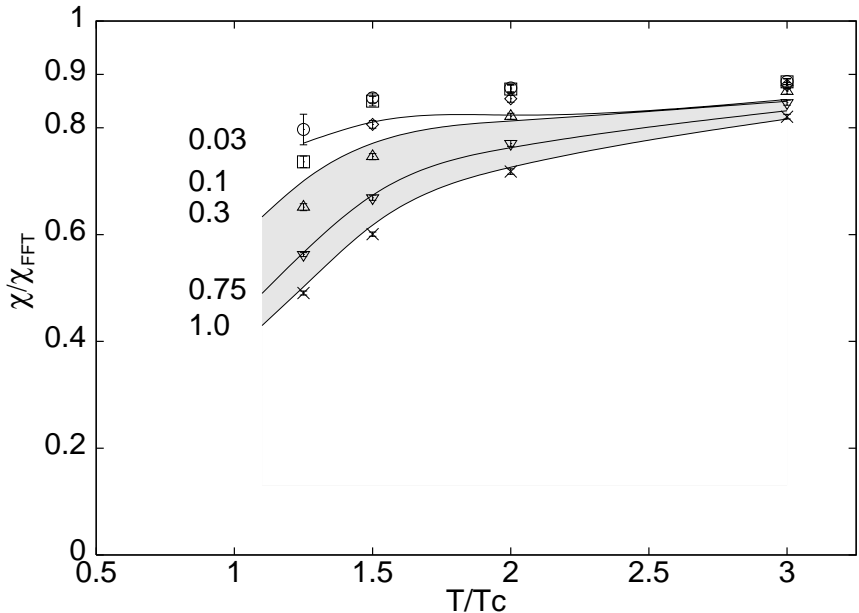}
\vspace{-0.5cm}
\caption{$\chi/\chi_{FFT}$ as a function of temperature for various
valence quark masses.}
\vspace{-0.5cm}
\label{fg.sus}
\end{center}
\end{figure}

Based on our tests \cite{qsus} of volume dependence, made by varying $N_s$ from
8 to 16, we chose $N_s =12$, as the susceptibilities differed very little from
those obtained on an $N_s=16$ lattice.  Choosing $N_t=4$, the temperature is $T
= (4a)^{-1}$, where the lattice spacing $a$ depends on the gauge coupling
$\beta$.  Using the known $\beta_c(N'_t)$ for $N'_t = 6, 8,
12$, where $\beta_c$ is the gauge coupling at which chiral (deconfinement)
transition/cross-over takes place for lattices with temporal extent $N'_t$,  we
obtained results at $T/T_c = N'_t/N_t$ =1.5, 2 and 3 in both the quenched
approximation and the 2-flavour QCD.  From the existing estimates \cite{tc} of
$T_c$ for 2-flavour QCD, one finds that the sea quark mass in our dynamical
simulations corresponds to 14-17 MeV.  Fig. \ref{fg.sus} shows our results,
normalized to the free field values on the same size lattice, $\chi_{FFT}$. 
These can be computed by setting gauge fields on all links to unity: $U_\nu(x)
=1$ for all $\nu$ and $x$.  Note that due this choice of our normalization, the
overall factor $N_f$ for degenerate flavours cancels out, permitting us to
exhibit both $\chi_3$ and $\chi_s$ of eq.(\ref{susc}) on the same scale in Fig.
\ref{fg.sus}.  The continuous lines in Fig. \ref{fg.sus} were obtained from the
interpolation of the data obtained in the quenched approximation (the data are
not shown for the sake of clarity), while the results for two light dynamical
flavours are shown by the data points.  In each case the value of $m_{val}/T_c$
is indicated on the left.  Due to the fact \cite{fsus} that the contribution of
${\cal O}_2$ to $\chi_s$ turns out to be negligibly small for $T > T_c$,
$\chi_3$ and $\chi_s$ appear coincident in Fig. \ref{fg.sus}.  For the real
world QCD, the low valence quark mass results are relevant for $\chi_3$ and
those for moderate valence quark masses are for $\chi_s$. 

Although $T_c$ differs in the quenched and 2-flavour QCD by a factor of
1.6-1.7, the respective susceptibilities shown in Fig. \ref{fg.sus} as a
function of the dimensionless variable $T/T_c$ change by at most 5-10\% for
any $m_{val}/T$.   Thus the effect of ``unquenching'', i.e., making 2 flavours
of quarks (u and d) light enough to include the contribution of the
corresponding quark loops, appears to be primarily a change of scale set by
$T_c$.  Since the strange quark is a lot heavier than the up and down quarks,
this suggests further that including its loop contributions, i.e., including a
dynamical but heavier strange quark, may not change the results in Fig.
\ref{fg.sus} significantly.  For a wide range for strange quark mass of 75 to
170 MeV, the strangeness susceptibility can be read off from the shaded region.
It is smaller by about 40\% compared to its ideal gas value near $T_c$ and the
suppression shows a strong temperature dependence.  This has implications for
the phenomenology of particle abundances, where an ideal gas model without such
a suppression of strangeness is employed and could therefore result in 
underestimates of the temperatures reached.  

\begin{figure}[htb]
\vspace{-0.5cm}
\begin{center}
\includegraphics*[angle=270,width=25pc]{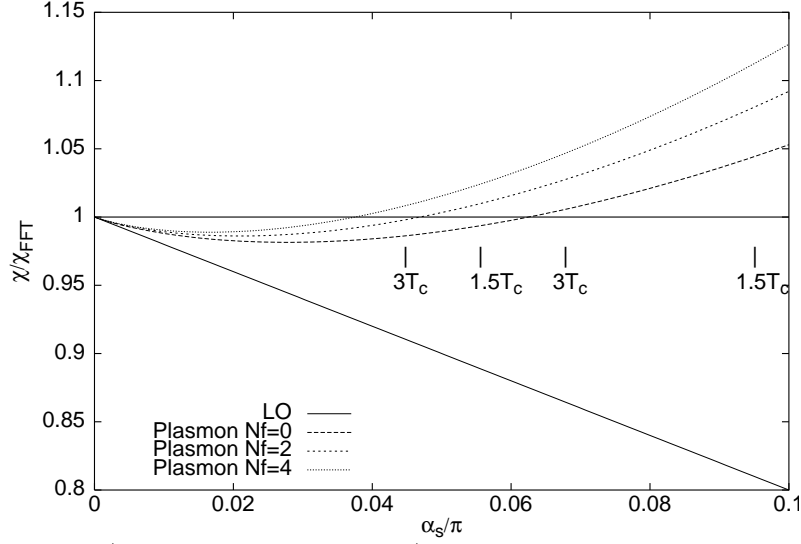}
\vspace{-0.5cm}
\caption{$\chi/\chi_{FFT}$ as a function of $\alpha_s/\pi$ for $N_f$ dynamical
massless quarks.}
\vspace{-0.5cm}
\label{fg.pert}
\end{center}
\end{figure}

For the smallest $m_{val}/T$ and highest temperature we studied, the ratio
$\chi/\chi_{FFT}$ in Fig. \ref{fg.sus} is seen to be 0.88 (0.85) for 2-flavour
(quenched) QCD, with a mild temperature variation in the large $T$-region.
Since its variation with valence quark mass is negligibly small for small
$m_{val}/T$, one can assume the results for massless valence quarks to be
essentially the same as those for $m_{val}/T =0.03$ in Fig. \ref{fg.sus}.  In
order to check whether the degrees of freedom of QGP can really be counted
using these susceptibilities,  one needs to know whether the deviation from
unity can be explained in ordinary perturbation theory or its improved/resummed
versions.  Usual weak coupling expansion \cite{jk} yields $\chi/\chi_{FFT} = 1
- 2{\alpha_s \over \pi} [1 - 4 \sqrt{{\alpha_s \over \pi}(1 + {N_f \over
  6})}]$.  Fig. \ref{fg.pert} shows these predictions for various $N_f$ along
with the leading order $N_f$-independent prediction.  Using a scale $2\pi T$
for the running coupling and $T_c/\Lambda_{\overline{MS}} = 0.49 (1.15)$ for
the $N_f = 2(0)$ theory \cite{tc}, the values $T/T_c$ =1.5 and 3 are marked on
the figure as the second (first) set.  As one can read off from the Fig.
\ref{fg.pert}, the ratio decreases with temperature in {\it both} the cases in
the range up to 3$T_c$ whereas our results in Fig.  \ref{fg.sus} display an
increase.  Furthermore, the perturbative results 
\begin{figure}[htb]
\vspace{-0.5cm}
\begin{center}
\includegraphics*[width=28pc]{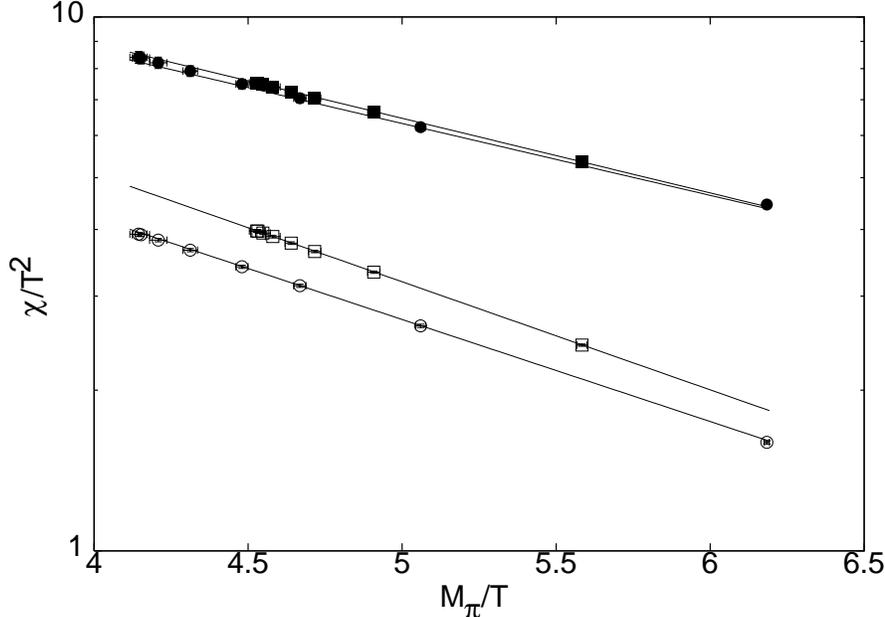}
\vspace{-0.5cm}
\caption{$4\chi_3/T^2$ (open symbols) and $\chi_\pi/10T^2$ (filled symbols) 
as a function of $M_\pi/T$ at $2T_c$ (circles) and $3T_c$ (boxes).}
\vspace{-0.5cm}
\label{fg.pi}
\end{center}
\end{figure}
lie significantly above in
each case, being in the range 1.027--1.08 for 2-flavour QCD and 0.986--0.994
for quenched QCD.  Although the order of magnitude of the degrees of freedom
can be gauged from these results and their eventual comparison with
experiments, they do call for clever resummations of perturbation theory for a
more convincing and precise count.

Alternatively, the deviations from free field theory could stem from
non-perturbative physics.  One known indicator of non-perturbative physics in
the plasma phase is the screening length in the channel with quantum numbers of
pion.  While it exhibits chiral symmetry restoration above $T_c$ by being
degenerate with the corresponding scalar screening length, its value is known
to be much smaller than the free field value unlike that for other screening
lengths.  Fig.  \ref{fg.pi} shows our results for $\chi_3$ and $\chi_\pi$
(defined as a sum of the pion correlator over the entire lattice) as a function
of the inverse pionic screening length, $M_\pi/T$.  It suggests the
non-perturbative physics in the two cases to be closely related, if not
identical.

\section{Acknowledgments}
It is a pleasure to thank my collaborators Sourendu Gupta and Pushan Majumdar.
I am also thankful to the Alexander von Humboldt Foundation for organizing
this wonderful symposium in its characteristic perfect style. It is a delight to
acknowledge the warm hospitality of the Physics Department of the University of
Bielefeld, especially that from Profs. Frithjof Karsch and Helmut Satz.


\begin{thebibliography}{9}
\bibitem{qsus} R. V. Gavai and S. Gupta, Phys. Rev. {\bf D 64} (2001) 074506.
\bibitem{fsus} Rajiv V. Gavai, Sourendu Gupta and Pushan Majumdar,
{\tt hep-lat/0110032}, Phys. Rev. {\bf D}, in press. 
\bibitem{satz} H. Satz, in these proceedings.
\bibitem{creutz} M. Creutz, {\sl ``Quarks, Gluons and Lattices''\/}, 1985,
      Cambridge University Press. 
\bibitem{bm} M. Asakawa, U. W. Heinz, and B. M\"uller, Phys. Rev. Lett. 
{\bf 85} (2000) 2072.
\bibitem{vk} S. Jeon and V. Koch, Phys. Rev. Lett. {\bf 85} (2000) 2076.
\bibitem{jrbm}J. Rafelski and B. M\"uller, Phys. Rev. Lett. {\bf 46} 
(1982) 1066; erratum-ibid {\bf 56} (1986) 2334.
\bibitem{resum} 
   J. P. Blaizot {\sl et al.\/}, Phys. Rev. {\bf D 63} (2001) 065003;
   J. O. Andersen {\sl et al.\/}, Phys. Rev.{\bf  D 63} (2001) 105008;
   K. Kajantie {\sl et al.\/}, Phys. Rev. Lett. {\bf 86} (2001) 10.        
\bibitem{rly} 
 S. Gottlieb {\sl et al.\/}, Phys. Rev. Lett. {\bf 59} (1987) 1513; 
 R. V. Gavai {\sl et al.\/}, Phys. Rev. {\bf D 40} (1989) 2743;
 S. Gottlieb {\sl et al.\/}, Phys. Rev. {\bf D 55} (1997) 6852.
\bibitem{chem} 
 R. V. Gavai, Phys. Rev. {\bf D 32} (1985) 519.
\bibitem{tc} 
   A. Ali Khan {\sl et al.\/}, Phys. Rev. {\bf D 63} (2001) 034502;
   F. Karsch {\sl et al.\/}, Nucl. Phys. {\bf B 605} (2001) 579;            
   S. Gupta, Phys. Rev. {\bf D 64} (2001) 034507.
\bibitem{jk} 
   J. I. Kapusta, {\sl ``Finite-temperature Field Theory''\/}, 1989,
      Cambridge University Press, Cambridge, UK, pp 132-133.     
\end{thebibliography}
\end{document}